\documentclass[twocolumn,showpacs]{revtex4b3}

\usepackage{graphicx}
\usepackage{dcolumn}
\usepackage{amsmath}

\def\TL{\hfil$\displaystyle{##}$}
\def\TR{$\displaystyle{{}##}$\hfil}
\def\TC{\hfil$\displaystyle{##}$\hfil}
\def\TT{\hbox{##}}
\def\seqalign#1#2{\vcenter{\openup1\jot
  \halign{\strut #1\cr #2 \cr}}}

\def\comment#1{}
\def\fixit#1{}

\def\mop#1{\mathop{\rm #1}\nolimits}

\def\overleftrightarrow#1{\vbox{\ialign{##\crcr
     $\leftrightarrow$\crcr\noalign{\kern-0pt\nointerlineskip}
     $\hfil\displaystyle{#1}\hfil$\crcr}}}

\def\lsim{\mathrel{\mathstrut\smash{\ooalign{\raise2.5pt\hbox{$<$}\cr\lower2.5pt\hbox{$\sim$}}}}}
\def\gsim{\mathrel{\mathstrut\smash{\ooalign{\raise2.5pt\hbox{$>$}\cr\lower2.5pt\hbox{$\sim$}}}}}
\def\sqr#1#2{{\vcenter{\vbox{\hrule height.#2pt
         \hbox{\vrule width.#2pt height#1pt \kern#1pt
            \vrule width.#2pt}
         \hrule height.#2pt}}}}
\def\square{\mathop{\mathchoice\sqr56\sqr56\sqr{3.75}4\sqr34\,}\nolimits}
\def\href#1#2{#2}  

\def\lbldef#1#2{\expandafter\gdef\csname #1\endcsname {#2}}
\def\eqn#1#2{\lbldef{#1}{(\ref{#1})}%
\begin{equation} #2 \label{#1} \end{equation}}
\def\eqalign#1{\vcenter{\openup1\jot
    \halign{\strut\span\TL & \span\TR\cr #1 \cr
   }}}

\def\Re{\mop{Re}}
\def\Im{\mop{Im}}

\begin{document}

\begin{widetext}\begin{flushright}
PUPT-1953 \\ hep-th/0009126
\end{flushright}\end{widetext}

\title[Short Title]{Instability of charged black holes \\ 
 in anti-de Sitter space}

\author{S. S. Gubser}
 \email{ssgubser@viper.princeton.edu}
\author{I. Mitra}
 \email{imitra@princeton.edu}
\affiliation{Joseph Henry Laboratories, Princeton University, Princeton,
NJ 08544}

\date{September, 2000}

\begin{abstract}
 We exhibit a tachyonic mode in a linearized analysis of perturbations
of large anti-de Sitter Reissner-Nordstrom black holes in four
dimensions.  In the large black hole limit, and up to a $0.7\%$
discrepancy which is probably round-off error in the numerical
analysis, the tachyon appears precisely when the black hole becomes
thermodynamically unstable.

\end{abstract}

\pacs{04.70.Bw, 11.25.Mj, 04.65.+e}

\maketitle

%\tableofcontents

%--------+---------+---------+---------+---------+---------+---------+
%Body
\section{Introduction}
\label{Introduction}

A curious dichotomy in the physics of black holes is that they
typically are thermodynamically unstable, in that they have negative
specific heat; but in a classical treatment, they are stable against
small perturbations of the metric (see for example
\cite{WaldSch,PressTeukolsky} for discussions of Schwarzschild and
Kerr black holes in asymptotically flat space).  In the past several
years, following \cite{StromingerVafa}, there has been remarkable
progress in providing a microscopic, statistical mechanical account of
black hole thermodynamics using string theory constructions such as
intersecting D-branes (for a recent pedagogical review, see
\cite{PeetTASI}).  The black holes so described without exception have
positive specific heat.  Typically, they are near-extremal solutions
to four- or five-dimensional compactifications of string theory, with
several electric and/or magnetic charges and a mass which almost
saturates the BPS bound.  The statistical mechanical account of their
entropy relies on a low-energy field theory description of the
D-branes from which they are constructed.  It is no surprise, then,
that the specific heat turns out to be positive: this is a criterion
which is met by the statistical mechanics of almost any sensible field
theory.

In the search for ways to extend the string theory's successes to more
astrophysically relevant black holes, a natural first step is to
search for thermodynamically unstable variants of the black holes for
which string theory provides a dual description.  In this Letter, we
exhibit perhaps the simplest example of such a black hole (the anti-de
Sitter space Reissner-Nordstrom solution), demonstrate its
thermodynamic instability, and show via numerics that the solution is
{\it unstable} in a linearized analysis.  The instability should
correspond to the onset of Bose condensation in the dual field theory
\cite{gspin}.  We conjecture a general link between thermodynamic
instability for black branes and the Gregory-Laflamme instability
\cite{glOne}.$^{\footnote{A recent paper \cite{Prestidge} has
investigated unstable modes of black hole solutions which owe their
existence to the non-trivial topology of periodic Euclidean time.  A
connection of these modes to thermodynamic instability was conjectured
in \cite{HawkingPage} and born out by the calculations of
\cite{Prestidge}.  The current work is somewhat different in that we
examine classical dynamical stability in Lorentzian signature, subject
to conservation of total mass and charge of the black hole.}}$ A
fuller treatment will appear in \cite{smForth}.

\section{The $AdS_4$-RN solution and its thermodynamics}
\label{Thermodynamics}

The anti-de Sitter space Reissner-Nordstrom solution ($AdS_4$-RN) is
  \eqn{AdSRN}{\seqalign{\span\TC}{
   ds^2 = -f dt^2 + {dr^2 \over f} + r^2 d\Omega^2 \qquad
   F_{0r} = {Q \over \sqrt{8} r^2}  \cr
   f = 1 - {2M \over r} + {Q^2 \over r^2} + {r^2 \over L^2} \,.
  }}
 We will work throughout in units where $G_4=1$.  The solution~\AdSRN\
can be embedded into M-theory in the following way.  A large number of
coincident M2-branes in eleven dimensions have as their near-horizon
geometry $AdS_4 \times S^7$.  There are eight dimensions transverse to
the M2-branes, and hence four independent angular momenta which the
branes can acquire if they are near-extremal.  If all four angular
momenta are equal, the resulting solution is a warped product of
$AdS_4$-RN and a deformed $S^7$.  Details can be found in
\cite{cejm}.$^{\footnote{We are glossing over a subtlety, namely that
the reduction on $S^7$ of the spinning, near-extremal M2-brane
solution in asymptotically flat eleven-dimensional spacetime is the
black brane limit of \AdSRN.  A solution to eleven-dimensional
supergravity which is asymptotically $AdS_4 \times S^7$ can be found
whose reduction is precisely \AdSRN.}}$  In fact, $AdS_4$-RN is a
solution of ${\cal N}=8$ gauged supergravity, whose maximally
supersymmetric $AdS_4$ vacuum is the Kaluza-Klein reduction of the
$AdS_4 \times S^7$ vacuum of M-theory.  Furthermore, ${\cal N}=8$
gauged supergravity is a consistent truncation of eleven-dimensional
supergravity \cite{deWitThree}, which means that any classical
solution of the four-dimensional theory lifts to an exact classical
solution of the eleven-dimensional theory.  Thus any instability found
in four dimensions is guaranteed to persist in eleven.

The lagrangian of ${\cal N}=8$ gauged supergravity contains the terms
  \eqn{NEightL}{\eqalign{
   &{\cal L} = {\sqrt{g} \over 16\pi} \Bigg[ R - 
    \sum_{i=1}^3 \left( {1 \over 2} (\partial\varphi_i)^2 + 
     {2 \over L^2} \cosh\varphi_i \right)  \cr
   &\qquad\quad - 
     2 \sum_{A=1}^4 e^{\alpha^A_i \varphi_i} 
     (F_{\mu\nu}^{(A)})^2 \Bigg]
  }}
 where
  \eqn{AlphaDef}{
   \alpha^A_i = \left( \begin{array}{cccc}
                           1 & 1 & -1 & -1  \\
                           1 & -1 & 1 & -1  \\
                           1 & -1 & -1 & 1
                       \end{array} \right)
  }
 and admits the black hole solutions \cite{DuffLiu}
  \eqn{DuffSoln}{\eqalign{
   ds^2 &= -{F \over \sqrt{H}} dt^2 + {\sqrt{H} \over F} dz^2 +
    \sqrt{H} z^2 d\Omega^2  \cr
   e^{2\varphi_1} &= {h_1 h_2 \over h_3 h_4} \quad
   e^{2\varphi_2} = {h_1 h_3 \over h_2 h_4} \quad
   e^{2\varphi_1} = {h_1 h_4 \over h_2 h_3}  \cr
   F_{0z}^{(A)} &= \pm {1 \over \sqrt{8} h_A^2} {Q_A \over z^2} \cr
   H &= \prod_{A=1}^4 h_A \quad 
   F = 1 - {\mu \over z} + {z^2 \over L^2} H \quad
   h_A = 1 + {q_A \over z}  \cr
   Q_A &= \mu \cosh\beta_A \sinh\beta_A \quad
   q_A = \mu \sinh^2 \beta_A
  }}
 for which the mass and entropy are
    \eqn{MassDef}{
   M = {\mu \over 2} + {1 \over 4} \sum_{A=1}^4 q_A  \qquad
   S = \pi z_H^2 \sqrt{H(z_H)} \,,
  }
  where $z_H$ is the largest root of $F(z_H) = 0$.  Only for a certain
range of the parameters $(\mu,q_A)$ do roots to this equation exist at
all.  When they don't, the solution is nakedly singular.  The
conserved physical charges are the $Q_A$, and they correspond to the
four independent angular momenta of M2-branes in eleven dimensions.
If all the $Q_A$ have a common value, $Q$, the solution \DuffSoln\
reduces to \AdSRN\ upon the change of variable $r=z+q$.

In the AdS/CFT correspondence \cite{juanAdS,gkPol,witHolOne} (see
\cite{MAGOO} for a review), it is claimed that 11-dimensional
supergravity on $AdS_4 \times S^7$ is physically equivalent to the
large $N$ limit of a 2+1-dimensional supersymmetric conformal field
theory (CFT) which lives on the boundary of $AdS_4$ and represents the
low-energy limit of the world-volume dynamics of $N$ coincident
M2-branes.  The electric charges $Q_A$ become global $R$-symmetry
charges in the CFT.  The solutions \DuffSoln, for sufficiently large
$M$, correspond to thermal states in the CFT with chemical potentials
for the global charges turned on.  It was argued in \cite{gspin} (in
fact for the somewhat simpler case of electrically charged black holes
in $AdS_5$) that the dual description of the thermodynamic instability
that we will point out in the next paragraph is an instability toward
condensation of the bosons carrying the relevant $U(1)$ global
charge(s).

For simplicity, let us now set $Q_1=Q_3$ and $Q_2=Q_4$, and consider
only the limit of large black holes, $M/L \gg 1$.  As $M/L \to
\infty$, one obtains a black brane solution in the Poincar\'e patch of
$AdS_4$.  Formally, this limit can be taken by expanding \DuffSoln\ to
leading order in small $\beta_i$, dropping the $1$ from $F$, and
replacing ${\bf S}^2$ by ${\bf R}^2$ in the metric.  A simple expression for
the mass can now be given:
  \eqn{MForm}{
   M = {1 \over 2\pi L^2} 
     \sqrt{(S^2 + \pi^2 L^2 Q_1^2)(S^2 + \pi^2 L^2 Q_2^2)
    \over \pi S} \,,
  }
 with corrections suppressed by powers of $M/L$.  Local thermodynamic
instability can now be expressed as convexity of the function
$M(S,Q_1,Q_2)$.$^{\footnote{To quantify thermodynamic stability
completely, we should demand convexity of $M$ as a function of
$S,Q_1,Q_2,Q_3$, and $Q_4$.  This stricter convexity requirement fails
at the same time as the one dealt with in the main text for the
$AdS_4$-RN solution.}}$ By forming the Hessian of $M(S,Q_1,Q_2)$, it
is straightforward to verify that convexity fails along the line $Q_1
= Q_2 = Q$ when $\pi L Q > S$, or equivalently when $M \sqrt{L} <
Q^{3/2}$.  The associated eigenvector has the form $(0,1,-1)$: it
looks like one charge wants to increase while the other decreases.  Of
course, this can happen only locally on account of global charge
conservation.  The calculation described in this paragraph is a
special case of \cite{cgTwo}.  It is worth noting that a black hole
horizon exists in the large black hole limit if and only if $M
\sqrt{L} \geq {2 \over 3^{3/4}} Q^{3/2}$.  Thus there is a narrow
range of thermodynamically unstable $AdS_4$-RN black holes which
borders on solutions which are nakedly singular.

\section{The existence of a tachyon}
\label{Tachyon}

Although there is now a certain literature on thermodynamic
instabilities of spinning branes \cite{gspin,CaiSoh,cgOne,cgTwo}, the
question has been left completely open whether there is a dynamical
instability.  The existence of a field theory dual makes this seem
almost inescapable: if the instability indeed indicates the onset of
Bose condensation, shouldn't the condensing bosons tend to clump
together on account of the ``attraction'' of
statistics?$^{\footnote{We do not have any description, either in the
CFT or in supergravity, of the condensed state.  It is possible that
there is no stable state at all for large $Q/S$, but it is impossible
to draw definite conclusions on this point in an analysis that
explores only unstable perturbations around an unstable extremum.}}$
There now seems to be a conflict of intuitions: on one hand, AdS/CFT
suggests that there should be a dynamical clumping instability, while
on the other hand we know that black holes in flat space are stable
against small perturbations, and it appears sensible to extend this
expectation at least to small black holes in AdS.  Also, general
arguments based on the dominant energy condition have been advanced
\cite{Hawking,HawkingStrings} to show that electrically charged black
holes in AdS are stable.

The resolution we suggest is that thermodynamic arguments in
AdS/CFT only make sense for large black holes, that there is a
clumping instability, and that the arguments of
\cite{Hawking,HawkingStrings} don't apply because there are matter
fields which violate the dominant energy condition.  Specifically, the
scalar potential in \NEightL\ gives the scalars negative mass-squared,
so $T^{\mu\nu} \xi_\mu$ is not necessarily forward-directed timelike
when $\xi^\mu$ is.

Because the unstable eigenvector of the Hessian of $M(S,Q_1,Q_2)$ did
not involve a variation of $S$, it is sensible to think that the
unstable perturbation need not involve the metric: rather, it should
take the form
  \eqn{FVar}{
   F^{(A)} = F + \alpha^A_1 \delta F \,,
  }
 accompanied by some variation in the scalars.  The quantities
$\alpha^A_i$ were defined in \AlphaDef.  Note that $\delta F$ is not
the variation in the background field strength $F$ of the $AdS_4$-RN
solution; rather, $F^{(1)}$ and $F^{(2)}$ increase by $\delta F$ while
$F^{(3)}$ and $F^{(4)}$ decrease by the same amount.  The metric does
not couple to fluctuations of the form \FVar\ at linear order because
the stress tensor is invariant at this order: $\sum_A F^{(A)} \cdot
\delta F^{(A)} = 0$.

It is straightforward to start with the lagrangian in \NEightL\ and
show that linearized perturbations to the equations of motion result
in the following coupled equations:
  \eqn{CoupledScalar}{\eqalign{
   & d\delta F = 0 \qquad 
     d*\delta F + d\delta\varphi_1 \wedge *F = 0  \cr
   & \left[ \square + {2 \over L^2} - 8 F_{\mu\nu}^2 \right] 
      \delta\varphi_1 - 16 F^{\mu\nu} \delta F_{\mu\nu} = 0 \,.
  }}
 Here $\square = g^{\mu\nu} \nabla_\mu \partial_\nu$ is the usual
scalar laplacian.  Variations in the other scalars, $\delta\varphi_2$
and $\delta\varphi_3$, do not couple and may be consistently set to
zero.  It is possible to eliminate the gauge field from
\CoupledScalar, at the expense of making the scalar equation fourth
order.  Details of the derivation of this equation are postponed to
the appendix.  Assuming a separated ansatz $\delta\varphi_1 = \Re
\left\{ e^{-i\omega t} Y_{\ell m} \delta\tilde\varphi_1(r) \right\}$,
and defining dimensionless quantities
  \eqn{uChiDef}{\eqalign{
   &u = {r \over M^{1/3} L^{2/3}} \qquad
    \tilde\omega = {\omega L^{4/3} \over M^{1/3}}  \cr
   &\chi = {Q \over M^{2/3} L^{1/3}} \qquad
    \sigma = \left( {L \over M} \right)^{2/3}  \cr
   &\tilde{f} = \sigma - 
      {2 \over u} + {\chi^2 \over u^2} + u^2 \,,
  }}
 one obtains the equation
  \eqn{FourthForm}{\eqalign{
   & \left( {\tilde\omega^2 \over \tilde{f}} + 
     \partial_u \tilde{f} \partial_u - 
     \sigma {\ell (\ell+1) \over u^2} \right)
     u^3  \cr
   &\quad{} \cdot \left( {\tilde\omega^2 \over \tilde{f}} + 
     \partial_u \tilde{f} \partial_u - 
     \sigma {\ell (\ell+1) \over u^2} - 
      {2 \over u^3} + {4\chi^2 \over u^4} \right)
     u \delta\tilde\varphi_1 =  \cr
   & \qquad\quad 4 \chi^2 \left( {\tilde\omega^2 \over \tilde{f}} + 
     \partial_u \tilde{f} \partial_u \right) \delta\tilde\varphi_1 \,.
  }}
 The black brane limit, where the horizon is ${\bf R}^2$ rather than
${\bf S}^2$, is $\sigma = 0$.  Only in this limit should we trust
thermodynamic arguments completely.  Away from it, finite size effects
may make the link between thermodynamic and dynamical instability
imprecise.  At $\sigma=0$, a horizon exists only if $\chi \leq
\sqrt{3}/2^{2/3} = 1.091$, and a thermodynamic instability appears for
$\chi > 1$.  Using Mathematica, we have found that there is a single
tachyonic mode in the $\delta\varphi_1$ equation for $\chi > 1.007$:
this is a normalizable solution with $\tilde\omega^2 < 0$.  The radial
wavefunction for $\delta\varphi_1$ falls off as $(r-r_H)^{|\omega|/f'(r_H)}$
near the black hole horizon, and as $1/r^2$ near the boundary of
$AdS_4$, indicating that the mode is indeed normalizable and
corresponds to a fluctuation in the state of the CFT rather than in
its lagrangian.  Thus, the range in which $AdS_4$-RN black holes in
the black brane limit are unstable compares fairly well with AdS/CFT
predictions: it is
  \eqn{SumUp}{\seqalign{\hfil\span\TT\quad & \span\TL & \span\TR}{
   AdS/CFT prediction: & 1 &< \chi \leq 1.091  \cr
   numerics: & 1.007 &< \chi \leq 1.091 \,.
  }}
 The small discrepancy is probably numerical error.  

 \begin{figure}
  \includegraphics{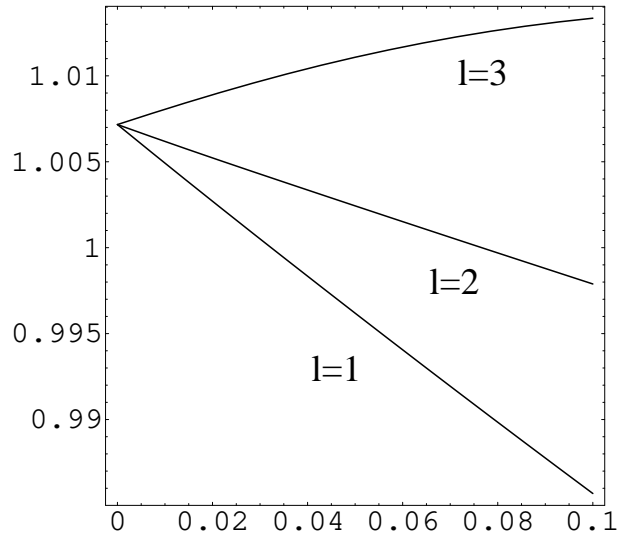}
  \caption{The boundaries at which various partial wave numbers begin
to exhibit tachyons.  The vertical axis is $\chi$ and the horizontal
axis is $\sigma$.}\label{figA}
 \end{figure} 
 The single tachyonic mode persists in each partial wave mode of
$\delta\varphi_1$ away from the strict black brane limit.  In
figure~\ref{figA} we exhibit a plot of the lines where tachyon
develops in the $\chi$-$\sigma$ plane for the first three partial
waves ($\ell=0$ is disallowed because of global charge conservation).
This tachyon is an example of the Gregory-Laflamme instability
\cite{glOne}.  In fact, it is to the authors' knowledge the first
proven example in a black hole geometry with a pointlike singularity
and a compact horizon.

\section{Remarks on the Gregory-Laflamme instability}
\label{Conclusion}

The Gregory-Laflamme instability is believed to be quite a general
phenomenon.  Though initially studied for black brane solutions which
were either the Schwarzschild solution cross ${\bf R}^p$ for some $p$
\cite{glOne}, or for charged $p$-branes which were far from
extremality in the sense that the mass was many times the BPS bound
\cite{glTwo}, it has been invoked in a variety of contexts (see for
example \cite{HorowitzMartinec,bdhm,chr,gregory}) when a black hole
horizon is thought to be unstable.  The intuitive explanation
\cite{glOne} for the instability is a thermodynamic one: the entropy
of an array of black holes is higher for a given mass than the entropy
of the uniform black brane.  This explanation leaves something to be
desired, since it applies equally to near-extremal D$p$-branes: a
sparse array of large black holes threaded by an extremal D$p$-brane
will be entropically favored over a uniformly non-extremal
D$p$-brane; however it is not expected that near-extremal D$p$-branes
exhibit the type of tachyonic mode found in \cite{glOne}.  Indeed, the
absence of tachyons in the extensive AdS-glueball calculations
(\cite{witHolTwo,Ooguri} and subsequent works---see \cite{MAGOO} for a
review) as provisional evidence that the static near-extremal
D3-brane, M2-brane, and M5-brane are (locally) stable.

We conjecture that {\it for a black brane with translational symmetry,
a Gregory-Laflamme instability exists precisely when the brane is
thermodynamically unstable.}  Here, by ``Gregory-Laflamme
instability'' we mean a tachyonic mode in small perturbations of the
horizon; and by ``thermodynamically unstable'' we mean that the
Hessian matrix of second derivatives of the mass with respect to the
entropy and the conserved charges or angular momenta has a negative
eigenvalue.  This conjecture fits the facts, in that the instabilities
observed in \cite{glOne,glTwo} are for black branes with negative
specific heat, whereas the AdS-glueball calculations were (mainly)
performed on thermodynamically stable black branes.  It also fits with
the predictions of AdS/CFT as far as we understand them.  The
stipulation of translational symmetry is clearly necessary to avoid
predicting the instability of the Schwarzschild solution in flat
space.

It may be possible to construct a general proof of our conjecture
along the following lines.  The supergravity action for the Euclidean
solution, when suitably regulated, is the free energy of the black
brane divided by the temperature.  Thermodynamic instability means
that this function lacks suitable convexity properties.  This is a
condition on the second variation of the (regulated) supergravity
action functional, and so might translate into an existence proof of
an unstable mode.  The main difficulty appears to be demonstrating
normalizability.

It should be possible to form a black hole very close to the
$AdS_4$-RN solution from nearly spherical collapse starting from
smooth initial conditions lying in some open set of configuration
space.  Since $AdS_4$-RN is unstable for a finite range of
parameters, and the Gregory-Laflamme instability is generically
expected to evolve into nakedly singular geometries, our results hint
that it might be possible to evolve smooth initial conditions into a
naked singularity in an asymptotically anti-de Sitter spacetime.

\section*{Acknowledgements}

We thank C.~Callan for useful discussions and H.~Reall for comments on
an early version of the manuscript.  This work was supported in part
by DOE grant~DE-FG02-91ER40671, and by a DOE Outstanding Junior
Investigator award.  S.S.G.\ thanks the Aspen Center for Physics for
hospitality during the early phases of the project.

\appendix

\begin{widetext}

\section{}

Decoupling the equations in \CoupledScalar\ is a chore greatly
facilitated by the use of the dyadic index formalism introduced in
\cite{NP}.  Making the definitions
  \eqn{Setup}{\eqalign{
   &l^\mu = (1/f,1,0,0) \quad 
    n^\mu = {1 \over 2} (1,-f,0,0) \quad
   m^\mu = {1 \over r \sqrt{2}} (0,0,1,i \csc\theta) \quad
    \bar{m}^\mu = {1 \over r \sqrt{2}} (0,0,1,-i \csc\theta)  \cr
   & \sigma^\mu_{\Delta\dot\Delta} = 
    \left( 
     \begin{array}{cc} l^\mu & m^\mu \\ \bar{m}^\mu & n^\mu \end{array} 
    \right) \qquad
    \sigma^\mu_{\Delta\dot\Delta} \partial_\mu = 
    \left(
     \begin{array}{cc} D & \delta \\ \bar\delta & \Delta \end{array}
    \right) \qquad
   \rho = -{1 \over r} \quad \mu = -{f \over 2r} \quad
    \gamma = {f' \over 4} \quad 
    \alpha = -\beta = -{\cot\theta \over \sqrt{8} r}  \cr
   & 4\sqrt{2}
   F_{\mu\nu} \sigma^\mu_{\Delta\dot\Delta} \sigma^\nu_{\Gamma\dot\Gamma} = 
    \Phi^{(0)}_{\Delta\Gamma} \epsilon_{\dot\Delta\dot\Gamma} +
    \bar\Phi^{(0)}_{\dot\Delta\dot\Gamma} \epsilon_{\Delta\Gamma} 
    \qquad
   4\sqrt{2} \delta F_{\mu\nu} 
    \sigma^\mu_{\Delta\dot\Delta} \sigma^\nu_{\Gamma\dot\Gamma} = 
    \Phi_{\Delta\Gamma} \epsilon_{\dot\Delta\dot\Gamma} +
    \bar\Phi_{\dot\Delta\dot\Gamma} \epsilon_{\Delta\Gamma}  \cr
   &\Phi^{(0)}_{\Delta\Gamma} = 
     \left( \begin{array}{cc} \phi_0^{(0)} & \phi_1^{(0)} \\
                \phi_1^{(0)} & \phi_2^{(0)}  \end{array} \right) \qquad
    \Phi_{\Delta\Gamma} =
     \left( \begin{array}{cc} \phi_0 & \phi_1 \\
                \phi_1 & \phi_2  \end{array} \right) \qquad
   \delta\varphi_1 = \varphi \,,
  }}
 one can cast the equations for $\delta\varphi_1$ and $\delta F$ into
the form
  \eqn{FirstPass}{\eqalign{
   &(D-2\rho) \phi_1 - (\bar\delta-2\alpha) \phi_0 =
     -\phi_1^{(0)} D\varphi  \qquad
   (\Delta+\mu-2\gamma) \phi_0 - \delta\phi_1 = 0  \cr
   &(D-\rho) \phi_2 - \bar\delta\phi_1 = 0  \qquad
   (\delta+2\beta) \phi_2 - (\Delta+2\mu) \phi_1 
     \phi_1^{(0)} \Delta\varphi  \cr
   &\left[ \square + {2 \over L^2} + 2 (\phi_1^{(0)})^2 \right] \varphi =
    -4 \phi_1^{(0)} \Re\phi_1 \,.
  }}
 Using the techniques of \cite{Teukolsky}, these equations can be
simplified to
  \eqn{SecondPass}{\eqalign{
   \left[ (D-3\rho)(\Delta+\mu-2\gamma) - 
    \delta(\bar\delta-2\alpha) \right] \phi_0 &= 
    -\phi_1^{(0)} \delta D\varphi  \cr
   \left[ (\Delta+3\mu)(D-\rho) - 
    \bar\delta(\delta+2\beta) \right] \phi_2 &=
    -\phi_1^{(0)} \bar\delta\Delta\varphi  \cr
   \left[ (D-2\rho)(\Delta+2\mu) - 
    (\delta+\beta-\alpha)\bar\delta \right] \phi_1 &= 
    -\phi_1^{(0)} D\Delta\varphi  \cr
   \left[ \square + {2 \over L^2} + 2 (\phi_1^{(0)})^2 \right] \varphi &=
    -4 \phi_1^{(0)} \Re\phi_1 \,,
  }}
 and then to 
  \eqn{GFODE}{
   \left[ (D-2\rho)(\Delta+2\mu) - (\delta+\beta-\alpha)\bar\delta \right]
    {1 \over 4\phi_1^{(0)}} \left[ \square + {2 \over L^2} + 
    2(\phi_1^{(0)})^2 \right] \varphi = \phi_1^{(0)} D\Delta\varphi \,.
  }
 In the last step we eliminated $\Re\phi_1$ algebraically using the
last equation in \SecondPass.  $\Im\phi_1$ decouples, and $\phi_0$ and
$\phi_2$ can be determined from \SecondPass\ once \GFODE\ is known.
These fields are also normalizable for the tachyon bound states found
in section~\ref{Tachyon}.  Using the separated ansatz $\varphi = \Re
\left\{ e^{-i\omega t} Y_{\ell m} \delta\tilde\varphi_1(r) \right\}$,
one obtains
  \eqn{FinalODE}{\eqalign{
   &\left( {\omega^2 \over f} + \partial_r f \partial_r - 
    {\ell (\ell+1) \over r^2} \right) r^3
    \left( {\omega^2 \over f} + \partial_r f \partial_r - 
    {\ell (\ell+1) \over r^2} - {2M \over r^3} + 
    {4 Q^2 \over r^4} \right) r \delta\tilde\varphi_1(r) =
     4 Q^2 \left( {\omega^2 \over f} + 
     \partial_r f \partial_r \right) \delta\tilde\varphi_1(r) \,,
  }}
 from which \FourthForm\ follows easily.

\vbox{\vskip0.5cm}

\end{widetext}

\bibliography{sum}

\begin{thebibliography}{10}
\expandafter\ifx\csname bibnamefont\endcsname\relax
  \def\bibnamefont#1{#1}\fi
\expandafter\ifx\csname bibfnamefont\endcsname\relax
  \def\bibfnamefont#1{#1}\fi
\expandafter\ifx\csname url\endcsname\relax
  \def\url#1{\texttt{#1}}\fi
\expandafter\ifx\csname urlprefix\endcsname\relax\def\urlprefix{URL }\fi
\providecommand{\bibinfo}[2]{#2}
\providecommand{\eprint}[2][]{\url{#2}}

\bibitem{WaldSch}
\bibinfo{author}{\bibfnamefont{R.~M.} \bibnamefont{Wald}}, \bibinfo{journal}{J.
  Math. Phys.} \textbf{\bibinfo{volume}{20}}, \bibinfo{pages}{1056}
  (\bibinfo{year}{1979}), \bibinfo{note}{erratum, {\bf 21} (1980) 218.}

\bibitem{PressTeukolsky}
\bibinfo{author}{\bibfnamefont{W.~H.} \bibnamefont{Press}} \bibnamefont{and}
  \bibinfo{author}{\bibfnamefont{S.~A.} \bibnamefont{Teukolsky}},
  \bibinfo{journal}{Atrophys. J.} \textbf{\bibinfo{volume}{185}},
  \bibinfo{pages}{649} (\bibinfo{year}{1973}).

\bibitem{StromingerVafa}
\bibinfo{author}{\bibfnamefont{A.}~\bibnamefont{Strominger}} \bibnamefont{and}
  \bibinfo{author}{\bibfnamefont{C.}~\bibnamefont{Vafa}},
  \bibinfo{journal}{Phys. Lett.} \textbf{\bibinfo{volume}{B379}},
  \bibinfo{pages}{99} (\bibinfo{year}{1996}), \eprint{hep-th/9601029}.

\bibitem{PeetTASI}
\bibinfo{author}{\bibfnamefont{A.~W.} \bibnamefont{Peet}}
  (\bibinfo{year}{2000}), \eprint{hep-th/0008241}.

\bibitem{gspin}
\bibinfo{author}{\bibfnamefont{S.~S.} \bibnamefont{Gubser}},
  \bibinfo{journal}{Nucl. Phys.} \textbf{\bibinfo{volume}{B551}},
  \bibinfo{pages}{667} (\bibinfo{year}{1999}), \eprint{hep-th/9810225}.

\bibitem{glOne}
\bibinfo{author}{\bibfnamefont{R.}~\bibnamefont{Gregory}} \bibnamefont{and}
  \bibinfo{author}{\bibfnamefont{R.}~\bibnamefont{Laflamme}},
  \bibinfo{journal}{Phys. Rev. Lett.} \textbf{\bibinfo{volume}{70}},
  \bibinfo{pages}{2837} (\bibinfo{year}{1993}), \eprint{hep-th/9301052}.

\bibitem{smForth}
 \bibinfo{note}{S. S. Gubser and I. Mitra, to appear.}

\bibitem{cejm}
\bibinfo{author}{\bibfnamefont{A.}~\bibnamefont{Chamblin}},
  \bibinfo{author}{\bibfnamefont{R.}~\bibnamefont{Emparan}},
  \bibinfo{author}{\bibfnamefont{C.~V.} \bibnamefont{Johnson}},
  \bibnamefont{and} \bibinfo{author}{\bibfnamefont{R.~C.} \bibnamefont{Myers}},
  \bibinfo{journal}{Phys. Rev.} \textbf{\bibinfo{volume}{D60}},
  \bibinfo{pages}{064018} (\bibinfo{year}{1999}), \eprint{hep-th/9902170}.

\bibitem{deWitThree}
\bibinfo{author}{\bibfnamefont{B.}~\bibnamefont{de~Wit}} \bibnamefont{and}
  \bibinfo{author}{\bibfnamefont{H.}~\bibnamefont{Nicolai}},
  \bibinfo{journal}{Nucl. Phys.} \textbf{\bibinfo{volume}{B281}},
  \bibinfo{pages}{211} (\bibinfo{year}{1987}).

\bibitem{DuffLiu}
\bibinfo{author}{\bibfnamefont{M.~J.} \bibnamefont{Duff}} \bibnamefont{and}
  \bibinfo{author}{\bibfnamefont{J.~T.} \bibnamefont{Liu}},
  \bibinfo{journal}{Nucl. Phys.} \textbf{\bibinfo{volume}{B554}},
  \bibinfo{pages}{237} (\bibinfo{year}{1999}), \eprint{hep-th/9901149}.

\bibitem{juanAdS}
\bibinfo{author}{\bibfnamefont{J.}~\bibnamefont{Maldacena}},
  \bibinfo{journal}{Adv. Theor. Math. Phys.} \textbf{\bibinfo{volume}{2}},
  \bibinfo{pages}{231} (\bibinfo{year}{1998}), \eprint{hep-th/9711200}.

\bibitem{gkPol}
\bibinfo{author}{\bibfnamefont{S.~S.} \bibnamefont{Gubser}},
  \bibinfo{author}{\bibfnamefont{I.~R.} \bibnamefont{Klebanov}},
  \bibnamefont{and} \bibinfo{author}{\bibfnamefont{A.~M.}
  \bibnamefont{Polyakov}}, \bibinfo{journal}{Phys. Lett.}
  \textbf{\bibinfo{volume}{B428}}, \bibinfo{pages}{105} (\bibinfo{year}{1998}),
  \eprint{hep-th/9802109}.

\bibitem{witHolOne}
\bibinfo{author}{\bibfnamefont{E.}~\bibnamefont{Witten}},
  \bibinfo{journal}{Adv. Theor. Math. Phys.} \textbf{\bibinfo{volume}{2}},
  \bibinfo{pages}{253} (\bibinfo{year}{1998}), \eprint{hep-th/9802150}.

\bibitem{MAGOO}
\bibinfo{author}{\bibfnamefont{O.}~\bibnamefont{Aharony}},
  \bibinfo{author}{\bibfnamefont{S.~S.} \bibnamefont{Gubser}},
  \bibinfo{author}{\bibfnamefont{J.}~\bibnamefont{Maldacena}},
  \bibinfo{author}{\bibfnamefont{H.}~\bibnamefont{Ooguri}}, \bibnamefont{and}
  \bibinfo{author}{\bibfnamefont{Y.}~\bibnamefont{Oz}}, \bibinfo{journal}{Phys.
  Rept.} \textbf{\bibinfo{volume}{323}}, \bibinfo{pages}{183}
  (\bibinfo{year}{2000}), \eprint{hep-th/9905111}.

\bibitem{cgTwo}
\bibinfo{author}{\bibfnamefont{M.}~\bibnamefont{Cvetic}} \bibnamefont{and}
  \bibinfo{author}{\bibfnamefont{S.~S.} \bibnamefont{Gubser}},
  \bibinfo{journal}{JHEP} \textbf{\bibinfo{volume}{07}}, \bibinfo{pages}{010}
  (\bibinfo{year}{1999}), \eprint{hep-th/9903132}.

\bibitem{cgOne}
\bibinfo{author}{\bibfnamefont{M.}~\bibnamefont{Cvetic}} \bibnamefont{and}
  \bibinfo{author}{\bibfnamefont{S.~S.} \bibnamefont{Gubser}},
  \bibinfo{journal}{JHEP} \textbf{\bibinfo{volume}{04}}, \bibinfo{pages}{024}
  (\bibinfo{year}{1999}), \eprint{hep-th/9902195}.

\bibitem{CaiSoh}
\bibinfo{author}{\bibfnamefont{R.-G.} \bibnamefont{Cai}} \bibnamefont{and}
  \bibinfo{author}{\bibfnamefont{K.-S.} \bibnamefont{Soh}},
  \bibinfo{journal}{Mod. Phys. Lett.} \textbf{\bibinfo{volume}{A14}},
  \bibinfo{pages}{1895} (\bibinfo{year}{1999}), \eprint{hep-th/9812121}.

\bibitem{Hawking}
\bibinfo{author}{\bibfnamefont{S.~W.} \bibnamefont{Hawking}} \bibnamefont{and}
  \bibinfo{author}{\bibfnamefont{H.~S.} \bibnamefont{Reall}},
  \bibinfo{journal}{Phys. Rev.} \textbf{\bibinfo{volume}{D61}},
  \bibinfo{pages}{024014} (\bibinfo{year}{2000}), \eprint{hep-th/9908109}.

\bibitem{HawkingStrings}
\bibinfo{note}{S. Hawking, ``Stability in ADS and Phase Transitions,'' talk at
  {\it Strings '99}, {\tt http://strings99.aei-potsdam.mpg.de\slash
  cgi-bin\slash viewit.cgi?speaker=Hawking}}.

\bibitem{glTwo}
\bibinfo{author}{\bibfnamefont{R.}~\bibnamefont{Gregory}} \bibnamefont{and}
  \bibinfo{author}{\bibfnamefont{R.}~\bibnamefont{Laflamme}},
  \bibinfo{journal}{Nucl. Phys.} \textbf{\bibinfo{volume}{B428}},
  \bibinfo{pages}{399} (\bibinfo{year}{1994}), \eprint{hep-th/9404071}.

\bibitem{HorowitzMartinec}
\bibinfo{author}{\bibfnamefont{G.~T.} \bibnamefont{Horowitz}} \bibnamefont{and}
  \bibinfo{author}{\bibfnamefont{E.~J.} \bibnamefont{Martinec}},
  \bibinfo{journal}{Phys. Rev.} \textbf{\bibinfo{volume}{D57}},
  \bibinfo{pages}{4935} (\bibinfo{year}{1998}), \eprint{hep-th/9710217}.

\bibitem{bdhm}
\bibinfo{author}{\bibfnamefont{T.}~\bibnamefont{Banks}},
  \bibinfo{author}{\bibfnamefont{M.~R.} \bibnamefont{Douglas}},
  \bibinfo{author}{\bibfnamefont{G.~T.} \bibnamefont{Horowitz}},
  \bibnamefont{and} \bibinfo{author}{\bibfnamefont{E.}~\bibnamefont{Martinec}}
  (\bibinfo{year}{1998}), \eprint{hep-th/9808016}.

\bibitem{chr}
\bibinfo{author}{\bibfnamefont{A.}~\bibnamefont{Chamblin}},
  \bibinfo{author}{\bibfnamefont{S.~W.} \bibnamefont{Hawking}},
  \bibnamefont{and} \bibinfo{author}{\bibfnamefont{H.~S.} \bibnamefont{Reall}},
  \bibinfo{journal}{Phys. Rev.} \textbf{\bibinfo{volume}{D61}},
  \bibinfo{pages}{065007} (\bibinfo{year}{2000}), \eprint{hep-th/9909205}.

\bibitem{gregory}
\bibinfo{author}{\bibfnamefont{R.}~\bibnamefont{Gregory}}
  (\bibinfo{year}{2000}), \eprint{hep-th/0004101}.

\bibitem{witHolTwo}
\bibinfo{author}{\bibfnamefont{E.}~\bibnamefont{Witten}},
  \bibinfo{journal}{Adv. Theor. Math. Phys.} \textbf{\bibinfo{volume}{2}},
  \bibinfo{pages}{505} (\bibinfo{year}{1998}), \eprint{hep-th/9803131}.

\bibitem{Ooguri}
\bibinfo{author}{\bibfnamefont{C.}~\bibnamefont{Csaki}},
  \bibinfo{author}{\bibfnamefont{H.}~\bibnamefont{Ooguri}},
  \bibinfo{author}{\bibfnamefont{Y.}~\bibnamefont{Oz}}, \bibnamefont{and}
  \bibinfo{author}{\bibfnamefont{J.}~\bibnamefont{Terning}},
  \bibinfo{journal}{JHEP} \textbf{\bibinfo{volume}{01}}, \bibinfo{pages}{017}
  (\bibinfo{year}{1999}), \eprint{hep-th/9806021}.

\bibitem{NP}
\bibinfo{author}{\bibfnamefont{E.}~\bibnamefont{Newman}} \bibnamefont{and}
  \bibinfo{author}{\bibfnamefont{R.}~\bibnamefont{Penrose}},
  \bibinfo{journal}{J. Math. Phys.} \textbf{\bibinfo{volume}{3}},
  \bibinfo{pages}{566} (\bibinfo{year}{1962}).

\bibitem{Teukolsky}
\bibinfo{author}{\bibfnamefont{S.~A.} \bibnamefont{Teukolsky}},
  \bibinfo{journal}{Astrophys. J.} \textbf{\bibinfo{volume}{185}},
  \bibinfo{pages}{635} (\bibinfo{year}{1973}).

\bibitem{Prestidge}
\bibinfo{author}{\bibfnamefont{T.}~\bibnamefont{Prestidge}},
  \bibinfo{journal}{Phys. Rev.} \textbf{\bibinfo{volume}{D61}},
  \bibinfo{pages}{084002} (\bibinfo{year}{2000}), \eprint{hep-th/9907163}.

\bibitem{HawkingPage}
\bibinfo{author}{\bibfnamefont{S.~W.} \bibnamefont{Hawking}} \bibnamefont{and}
  \bibinfo{author}{\bibfnamefont{D.~N.} \bibnamefont{Page}},
  \bibinfo{journal}{Commun. Math. Phys.} \textbf{\bibinfo{volume}{87}},
  \bibinfo{pages}{577} (\bibinfo{year}{1983}).

\end{thebibliography}

\end{document}